%% file: mnlong-pp-feb24a.tex
\documentstyle[psfig]{mn}

\newcommand{\ensuremath}[1]{#1}
\newcommand{\Tbb}{\ensuremath{T_{\mbox{\scriptsize\sc bb}}}}
\newcommand{\Abb}{\ensuremath{A_{\mbox{\scriptsize\sc bb}}}}
\newcommand{\Apl}{\ensuremath{A_{\mbox{\scriptsize\sc pl}}}}

\newcommand{\Ac}{\ensuremath{A_{\mbox{\scriptsize\sc c}}}}

\newcommand{\NH}{\ensuremath{N_{\mbox{\scriptsize\sc h}}}}

\newcommand{\Efold}{\ensuremath{E_{\mbox{\scriptsize\rm f}}}}
\newcommand{\taue}{\ensuremath{\tau_{\mbox{\scriptsize e}}}}
\newcommand{\Tc}{\ensuremath{T_{\mbox{\scriptsize\sc c}}}}

\newcommand{\Rc}{\ensuremath{R_{\mbox{\scriptsize\sc c}}}}
\newcommand{\errtwo}[3]{\ensuremath{#1^{+#2}_{-#3}}}
\newcommand{\colhead}[1]{\multicolumn{1}{c}{#1}}
\newcommand{\nodata}{\ldots}

\pssilent

\begin{document}

\title{RXTE Observation of Cygnus X-1: I. Spectral Analysis}

\author[J. B. Dove et al.]{James B. Dove$^{1,2}$\thanks{Present address:
  Department of Physics and Astronomy, University of Wyoming, Laramie, WY
  82071, USA}, J\"orn Wilms$^{3,1}$,
  Michael A. Nowak$^{1}$, \cr Brian A. Vaughan$^{4}$, and
  Mitchell C. Begelman$^{1,2}$\\
$^1$ JILA, University of Colorado
  and National Institute of Standards and Technology, Campus Box 440,
  Boulder, CO 80309-0440, USA\\
$^2$Department of Astrophysical and Planetary Sciences,
  University of Colorado, Boulder, CO 80309-0391, USA\\
$^3$Institut f\"ur Astronomie und Astrophysik,
  Abt.~Astronomie, Waldh\"auser Str. 64, D-72076 T\"ubingen, Germany\\
$^4$Space Radiation Laboratory, California Institute of
  Technology, 220-47 Downs, Pasadena, CA 91125, USA}

\maketitle

\begin{abstract}
  We present the results of the analysis of the broad-band spectrum of
  Cygnus~X-1 from 3.0 to 200\,keV, using data from a 10\,ksec observation
  by the {\it Rossi X-ray Timing Explorer}. The spectrum can be well
  described phenomenologically by an exponentially cut-off power law with a
  photon index $\Gamma = \errtwo{1.45}{0.01}{0.02}$ (a value considerably
  harder than typically found), e-folding energy $\Efold =
  \errtwo{162}{9}{8}$\,keV, plus a deviation from a power law that formally
  can be modeled as a thermal blackbody with temperature $k\Tbb
  =\errtwo{1.2}{0.0}{0.1}$\,keV.  
  Although the $3$--$30$ keV portion of the spectrum can be fit with a
  reflected power law with $\Gamma = 1.81 \pm 0.01$ and covering fraction
  $f = 0.35 \pm 0.02$, the quality of the fit is significantly reduced when
  the HEXTE data in the $30$--$100$ keV range is included, as there is no
  observed hardening in the power law within this energy range.
  As a physical description of this system, we apply the accretion disc
  corona models of Dove, Wilms \& Begelman \shortcite{dove:97a} --- where
  the temperature of the corona is determined self-consistently.  A
  spherical corona with a total optical depth $\tau = 1.6 \pm 0.1$ and an
  average temperature $k\Tc = 87\pm 5$\,keV, surrounded by an exterior cold
  disc, does provide a good description of the data ($\chi^2_{\rm red} =
  1.55$).  These models deviate from the data by up to 7\% in the $5 -
  10$\,keV range, and we discuss possible reasons for these
  discrepancies. However, considering how successfully the spherical corona
  reproduces the $10 - 200$\,keV data, such ``photon-starved'' coronal
  geometries seem very promising for explaining the accretion processes of
  Cygnus~X-1.
\end{abstract}


\setcounter{footnote}{0}

\section{Introduction}\label{sec:intro}
Cygnus~X-1 is one of the most firmly established persistent galactic black
hole candidates (BHCs). Since it is also one of the brightest BHCs, the
study of its X-ray spectrum can help reveal the physical conditions of the
inner region around accreting compact objects.  The X-ray spectrum of
Cyg~X-1 while in the hard (i.e., low) state can be roughly described by a
power-law with a photon-index $\Gamma \sim 1.65$, modified by an
exponential cutoff with an $\rm e$-folding energy $\Efold \sim 150$\,keV
(Ebisawa et al. 1996; Gierli\'nski et al. 1997, and references therein). A
spectral shape of this form is naturally explained by Comptonization of
low-energy seed photons by a semi-relativistic corona
\cite{sunyaev:79a,tit:94a}.  Below about 1\,keV, there is evidence for a
soft-excess, usually interpreted as thermal radiation from a cold accretion
disc having a temperature $k\Tbb \sim 0.1 - 0.3$\,keV
\cite{balu:95a,ebisawa96b}.  In addition, the spectrum is usually found to
contain a weak iron line feature at $\approx 6.4$\,keV and a slight
hardening above $10$\,keV, often interpreted as being due to Compton
reflection (Barr, White \& Page 1985; Ebisawa et al. 1996, and references
therein).

The combination of a Comptonization continuum and reprocessing features has
been interpreted as being due to an accretion disc corona (ADC).  The
geometric configuration of the corona and the cold disc, however, is still
unclear.  In most prior work, the geometry has been assumed to be a cold
accretion disc embedded between two hot coronae, with a slab geometry
(Haardt, Maraschi \& Ghisellini 1996; Haardt et al. 1993, and references
therein).  Recently, however, evidence has been presented showing that ADC
models with a slab geometry suffer from several problems, making them less
likely to be the appropriate models for explaining the high-energy
radiation of BHCs.  Gierli\'nski et al. \shortcite{gierlinski:97a} analyzed
simultaneous {\sl Ginga}-OSSE data of Cyg~X-1 and found that the strength
of the reflection component is too small to be consistent with a slab
geometry.  Using a non-simultaneous broad-band spectrum of Cyg~X-1, we
previously argued that our self-consistent ADC models with a slab geometry
(in which the coronal temperature is not a free parameter but is determined
by balancing Compton cooling with viscous dissipation) could not have a
high enough temperature\footnote{Formally, a high temperature is possible
for a slab corona if the coronal optical depth is very low.  However, such
low optical depth coronae also produce spectra that are too soft to explain
the observations of Cyg X-1.}  to explain both the observed power-law and
exponential cutoff \cite{dove:97a,dove:97b,wilms:97a}.  Moreover, the
luminosity of the disc expected due to reprocessing is comparable to the
coronal luminosity, and therefore the predicted soft-excess component is
much stronger than that observed \cite{dove:97b,gierlinski:97a}.

Most of the problems cited above are due to the cold disc of the slab ADC
model having a covering fraction of unity (i.e., all downward directed
coronal radiation is reprocessed in the cold disc).  ADCs having a geometry
with a smaller covering fraction have weaker reprocessing features, are
less efficiently Compton cooled, and thereby allow higher coronal
temperatures for a given $\tau$.  An example is a hot coronal sphere that is
surrounded by an exterior cold, optically thick, geometrically thin,
accretion disc (called the ``sphere+disc geometry'' henceforth), similar to
the two-temperature disc model of Shapiro, Lightman \& Eardley
\shortcite{shapiro:76a} and to advection dominated models
\cite{abramowicz:95a,chen:95a,narayan:94a}.  Based on non-simultaneous
BBXRT, Mir-TTM, Mir-HEXE, and OSSE data, we have shown that the sphere+disc
ADC model can explain the observed spectrum of Cyg~X-1 for energies in the 
$\sim 1$\,keV$ - 1$\,MeV range \cite{dove:97b}.

In this paper, we analyze the broad-band spectrum of Cyg~X-1 from 3.0 to
200\,keV, using data from the {\it Rossi X-ray Timing Explorer} (RXTE).
The simultaneity of this broad-band observation allows us to place stronger
constraints on the coronal parameters than was previously possible.  In
\S\ref{sec:obs}, we describe the observations performed and the
procedures used in the data analysis. In \S\ref{sec:model}, we give the
results from spectral modeling of the Cyg~X-1 spectrum using both
``standard'' simplified models as well as our self-consistent ADC models.
In \S\ref{sec:discuss} the results of these fits are discussed.

\section{Observations and Data Analysis}\label{sec:obs}
RXTE observed Cyg~X-1 for a total of 22.5\,ksec on 1996 October 23 and 24.
For the analysis presented here, we use data from the proportional counter
array (PCA) and the high energy X-ray timing experiment (HEXTE).  The PCA
consists of five Xe proportional counters with a total effective area of
about $6500\,{\rm cm}^2$ \cite{zhangw:93a}, while HEXTE consists of two
clusters of NaI(Tl)/CsI scintillation counters with a total effective area
of about $1400\,{\rm cm}^2$ \cite{rothschild:97a}.  For both instruments, only
data where the source was observed at elevations higher than 10$^\circ$
above the spacecraft horizon were used.  Due to missing standard-mode PCA
data, we have used only the first 10\,ksec of the observation.  This
duration is sufficient for the spectral analysis, as the PCA count-rate is
high ($\approx 4300$\,cps, with a background of $\approx 150$\,cps).

For the extraction of the PCA data, we used the standard RXTE ftools,
version 4.0.  We used version 1.5 of the background estimator program, in
which activation due to the South Atlantic Anomaly and an estimate for the
internal background based on the measurement of the rate of Very Large
Events within the PCA are taken into account. To this estimation, the
measured diffuse X-ray background is added by the program. The comparison
of the modeled background spectrum to the background measured by the PCA
during the Earth occultation phases of the observation (defined by an
elevation $<-5^\circ$) indicates that the background model count-rate has
an uncertainty of 5 to 10\% (i.e., $\la 0.3$\% of the total count rate).
The use of the background model is still preferable to using the Earth
occultation data because the latter method cannot estimate the variation of
the background due to the variation of the rigidity of the Earth's magnetic
field during on-source time intervals.

\begin{figure}
\centerline{\psfig{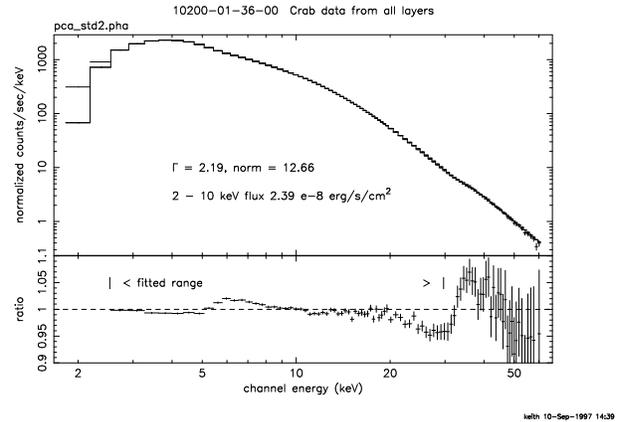}}
\caption{Top: Comparison of the Crab Pulsar spectrum, taken by PCA,
  to the best fit power law model. Bottom: Ratio of the residuals of the
  model to the data (data/model). Figure by K. Jahoda, private
  communication.}  \protect\label{fig:pca-crab}
\end{figure}

For the spectral analysis of the PCA we used version~2.2.1 of the PCA
response matrix (Jahoda, priv.\ comm.; for an overview of the PCA
calibration issues see Jahoda et al., 1996).  The PCA response matrix is
now fairly well understood.  An analysis of the residuals of a standard
power-law fit to the spectrum of the Crab pulsar indicates that the
uncertainty of the calibration in the vicinity of the Xenon L edge (around
$5.5$\,keV) is about $2\%$. This makes a precise determination of the
parameters of the Fe K$\alpha$ line difficult. In addition, the Crab fit
indicates a calibration uncertainty of about $5\%$ above $\approx 30$\,keV,
as shown in Figure~\ref{fig:pca-crab}.
Since HEXTE provides reliable data for energies $\ga 20$\,keV we ignored
the PCA data above 30\,keV.  To be conservative, we ignored the first four
PCA channels (i.e., we only included PCA data above 3\,keV).  To account
for the additional uncertainties in the PCA response matrix, a 2\%
systematic error was added to all PCA data.

\begin{figure}
\centerline{\psfig{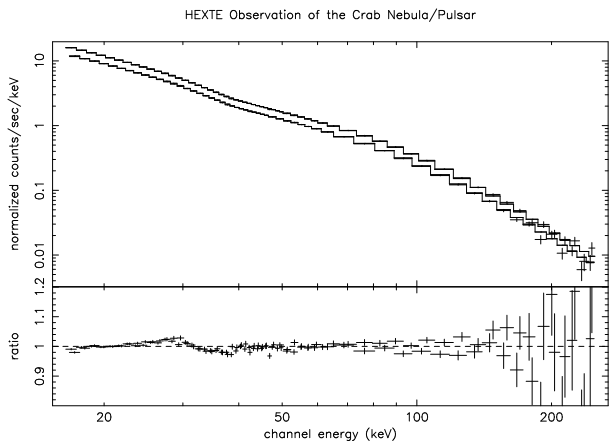}}
\caption{Top: Comparison of the Crab Pulsar spectrum, taken
  simultaneously by HEXTE Clusters A and B, to the best fit broken power
  law model ($\Gamma_1 = 2.045\pm 0.002$, $\Gamma_2 = 2.140\pm0.009$, and
  the break energy $E_B = 57.2\pm3.0$\,keV). Bottom: Ratio of the residuals
  of the model to the data (data/model) (Rothschild et al. 1997, Fig. 11).}
\protect\label{fig:hexte-crab}
\end{figure}

Software provided by the HEXTE instrument group was used for the extraction
of the HEXTE data and subsequent dead-time corrections.  Since HEXTE is
source-background swapping, the background is not a problem in the data
analysis.  We used the HEXTE response matrices released 1997 March 20.
Only data above 20\,keV were used due to the uncertainty of the response
matrix below these energies.  At high energies, the spectrum was cut at
200\,keV. To ensure good statistical accuracy above 50\,keV, the spectrum
was rebinned by a factor of 3 for channels between 50 and 100\,keV, and by
a factor of 10 for higher channels. The spectrum of the Crab Pulsar, taken
by HEXTE, is shown in Figure~\ref{fig:hexte-crab}.

\section{Spectral Analysis}\label{sec:model}
\subsection{Standard Models}\label{sec:classic}
Spectral fitting was first performed using the following standard models: a
power-law, a power-law with an exponential cutoff, a power-law with an
exponential cutoff plus a cold reflection component, and thermal
Comptonization models.  In addition, we added a Gaussian line (with energy
and width fixed to 6.4 keV and 0.1 keV, respectively) to several of these
models. We fixed the low-energy absorption to an equivalent cold Hydrogen
column of $\NH=6\times 10^{21}\,{\rm cm}^{-2}$, the value suggested by the
soft X-ray spectrum and interstellar reddening measurements of HDE~226868
\cite{balu:91a,wu:82a}.  The results of the spectral fits are given in
Table~\ref{tab1}\footnote{To account for the known discrepancies in the
relative normalization of PCA and HEXTE, we introduced a multiplicative
constant in the spectral models that represents the relative normalization
between the data-sets.  The relative normalization between HEXTE and PCA
was always found to be $0.74\pm0.01$ and is therefore not listed in
Table~\ref{tab1}.}.

\begin{figure}
\centerline{\psfig{file=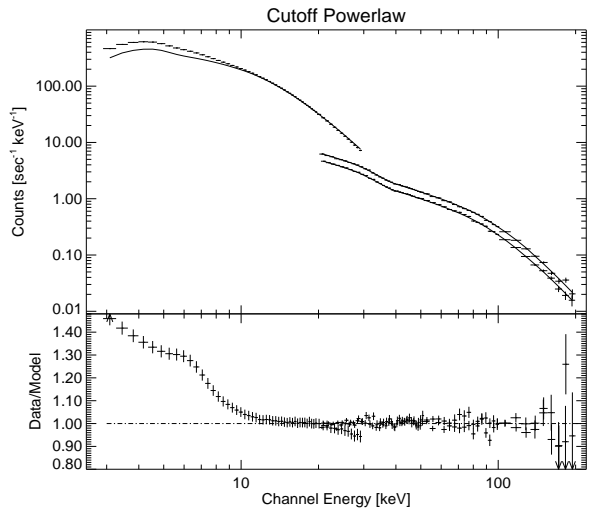,width=8cm,angle=0}}
\caption{Comparison of the RXTE spectrum of Cyg~X-1 to the best-fit
  power-law with an exponential cutoff model. This model was fit to only
  the $10$ - $200$\,keV data, yielding $\chi^2_{red} = 171/149$. The photon
  power-law index $\Gamma = 1.46$ and the $e$-folding energy $E_{\rm f} =
  168$\,keV. A soft-excess below $10$\,keV is clearly evident.}
\label{fig:cutoffpl-10-200}
\end{figure}

\begin{figure}
\centerline{\psfig{file=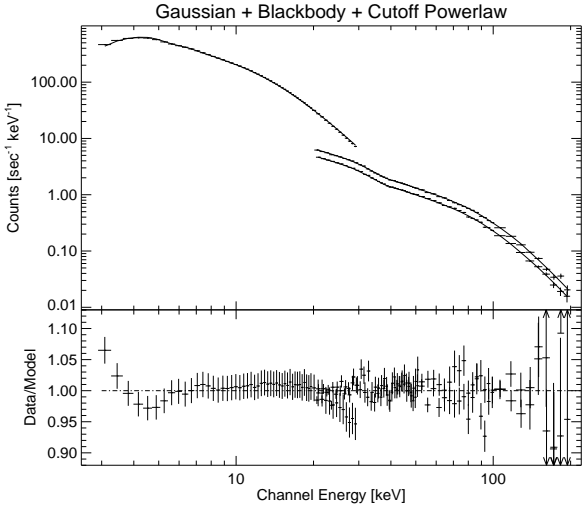,width=8cm,angle=0}}
\caption{Comparison of the RXTE spectrum of Cyg~X-1 to the best-fit
  power-law with an exponential cutoff model. The photon power-law index
  $\Gamma = 1.45$ and the $e$-folding energy is $E_{\rm f} = 164$\,keV. A
  blackbody component, with a temperature $kT_{\rm BB} = 1.1$\,keV and a
  relative flux of $\protect\la 5$\%, and a Gaussian component, with an energy
  fixed at $6.4$\,keV and an equivalent width EW$ = 46.4$\,eV, were added
  to fit the soft-excess. For this fit, $\chi^2_{\rm red} = 170/166$.}
\protect\label{fig:gauss_bbody_cutoffplb}
\end{figure}

For energies $E \ga 10$\,keV, a power-law with an exponential cutoff
provides a good description of the data, as also shown in
Figure~\ref{fig:cutoffpl-10-200}. 
The value for the e-folding energy, $\Efold$, is well constrained due to
the quality of the HEXTE data.  The analysis of the residuals of this fit
indicates the presence of a soft excess at energies $\la 8$\,keV, as shown
in Figure~\ref{fig:cutoffpl-10-200}. Adding a
black-body component with a temperature $k\Tbb \approx 1$\,keV and a
Gaussian component, with an equivalent width EW$= 46.4$\,eV, to the model
improves the quality of the fit, resulting in a good description of the
data over the whole energy range ($\chi^2_{\rm red}=170/166$), as shown in
Figure~\ref{fig:gauss_bbody_cutoffplb}.
We note that this black-body component should not be interpreted literally
as being due to disc emission; it is simply added phenomenologically to
measure the magnitude of the data's deviation from a power law. We could
have instead used a disc blackbody, Gaussian, or bremsstrahlung model with
roughly equal success. 

The addition of a reflection component to this model {\em does not} improve
the quality of the fit, as the best-fit value for the fraction of incident
radiation that is Compton reflected by cold matter is $f \la 0.02$.  Even
though we only considered PCA data in the $3.0$\,keV $ - $ $30$\,keV range,
the power-law index of the continuum is well constrained by the HEXTE data,
and there is no evidence that the spectral slope of the data for the two
instruments disagree within the energy range of the overlap. The PCA
residuals between 20 and 30 keV seen in Figures 3-6 are consistent with
those seen for power-law fits to the Crab, as shown in
Figure~\ref{fig:pca-crab}.

\begin{figure}
\centerline{\psfig{file=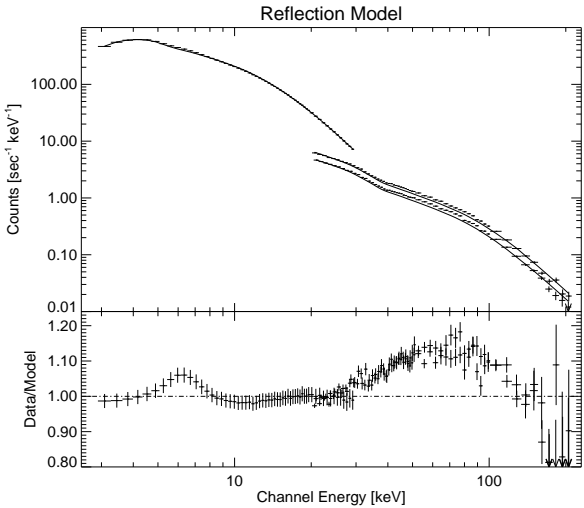,width=8.cm,angle=0.}}
\caption{Top: Comparison of the reflection model (Magdziarz \& Zdziarski
  1995) to the PCA data alone. Here,
  $\chi^2_{red} = 143/73$. Adding a Gaussian component, with a line energy
  $E_l = 6.4$\,keV, a line width $\Delta E = 64$\,eV, and an equivalent
  width $EW = 110$eV, results in a slightly better fit ($\chi^2_{red} =
  107/72$), and the other best-fit parameters do not change appreciably.
  Bottom: Ratio between data and model.}  \protect\label{fig:pexrav-3-30}
\end{figure}
We believe that the measured spectrum parameters, which represent a harder
power law and a weaker reflection component than previously found by
others, is mostly due to our fitting the entire $3 - 200$\,keV spectral
band.  Previous observations using different instruments have had
relatively poor spectral coverage in the energy range $\sim 30 -
100$\,keV. We do not believe that our results are heavily influenced by any
remaining uncertainties in the PCA response matrix.  To support this claim,
if we restrict the data to $3 - 30$\,keV and $100 - 200$\,keV (roughly
covering the energy range of the {\sl Ginga} + OSSE observations) we {\it
do} find best-fit parameter values similar to those previously found for
Cyg X-1.  Specifically, for this restricted data range, we find that a
reflected power law with $\Gamma = 1.81 \pm 0.01$ and covering fraction $f
= 0.35 \pm 0.02$ yields $\chi^2_{\rm red} = 157/90=1.75$.  Adding back the
HEXTE data in the $30 - 100$\,keV range, significantly reduces the quality
of the fit ($\chi^2_{\rm red} = 8.7$, $77$ dof), as shown in
Figure~\ref{fig:pexrav-3-30}.  Being able to fit a reflected power-law with
a large disc covering fraction is thus seen to be partly a consequence of
ignoring the data in the $\sim 30 - 100$\,keV range.

It is this $30 - 100$\,keV energy range that is crucial for constraining a
power-law index since these energies are uncontaminated by reprocessing or
reflection features, i.e., the final power law in this energy range is
identical to the intrinsic power law prior to reflection We therefore
postulate that the observed 10--30\,keV ``hardening'' feature typically
found by others appears to be the beginning of a hard power-law that
continues to $\sim 100$\,keV, and that the $\sim 1 - 10$\,keV portion of
the spectrum is softer due to contamination from the thermal excess. This
contamination could be due to Comptonization of the thermal radiation
emitted by the cold disc with an effective blackbody temperature $k\Tbb
\sim 150$\,eV.

We emphasize that this result does not indicate that there are no reflection
features in the observed spectrum of Cyg~X-1. As we discuss below, our ADC
model with a sphere+disc geometry is able to describe the observation, and
this model does include reflection.  The reflection feature predicted from
our model, due to reprocessing from the Comptonized continuum (which is not
a pure power-law due to the contribution from the seed photons, i.e.,
thermal emission of the disc), differs from the feature predicted by
phenomenological reflection models such as the popular {\em pexrav} model
\cite{magdziarz:95a}, in which a pure power law is incident onto a uniform
slab.

Although the residuals of our power-law fits clearly indicate the presence
of an Fe K$\alpha$ line, the $\sim 2$\% uncertainty in the PCA response
matrix at 6 keV make a determination of the line equivalent width
problematic.  The division of the Cyg~X-1 spectrum by a spectrum of the
Crab pulsar (which is assumed to be a pure power-law) also shows that the
Iron-line feature is present in our data. Due to different detector gains
between the two observations, however, we cannot use the divided spectrum
to estimate the equivalent width of the line.  We did include a Gaussian
line in most of our fits presented in Table~\ref{tab1}.  We fixed the line
energy at 6.4\,keV and the line-width to 100\,eV [which was the typical
value for the ASCA observations (Ebisawa et al. 1996); For all models,
thawing the line-width resulted in a best-fit width of $\sim
600$ eV.]. Fitting only the line amplitude, the fits typically yielded an
EW $\approx 60 \pm 35$\,eV. Although undoubtedly influenced by the
uncertainties in the response matrix, this EW does indicate a contribution
from an intrinsic line and is probably in agreement with the results based
on four days of ASCA observations presented by Ebisawa et al.
\shortcite{ebisawa96b}, who found a comparably weak Fe line with an
equivalent width EW$ \la 30$\,eV.

Thermal Comptonization is a more physical model for the observed energy
spectrum of Cyg~X-1.  Applying the model of Titarchuk \shortcite{tit:94a}
to the data, we find that a thermal Comptonization spectrum, resulting from
a spherical geometry with $\tau \approx 3.6$ and $k\Tc \approx 40$\,keV,
with an additional soft component, can give an acceptable description of
the data, although the model underestimates the flux at energies $\ga
150$\,keV.  The semi-relativistic, optically thin model \cite{tit:94a}, for
either a slab geometry or a spherical geometry was unable to explain the
data.

Although the traditional models can be reasonably successful in describing
the observed broad band spectrum of Cyg~X-1, they do not yield a physical
interpretation of the emitting mechanisms responsible for the production of
the high-energy radiation.  There need not be a one-to-one correspondence
between a phenomenological model component and a physical interpretation.
We therefore apply our self-consistent ADC models to the observed data.

\subsection{Accretion Disc Corona models}
For both a slab ADC model and the sphere+disc ADC model, we have computed
grids of spectra using a non-linear Monte Carlo scheme based on the code of
Stern et al. \shortcite{stern:95a} and modified by Dove et al. (1997a,b).
The free parameters of the model are the seed optical depth $\taue$, (the
optical depth of the corona excluding the contribution from
electron-positron pairs) and the heating rate (i.e., the compactness
parameter) of the ADC. For a given compactness parameter, the temperature
structure of the corona is determined by balancing Compton cooling with
heating, where the heating rate is assumed to be uniformly distributed.
The $e^-e^+$-pair opacity is given by balancing photon-photon pair
production with annihilation. Reprocessing of coronal radiation in the cold
accretion disc is also treated numerically
\cite{dove:97b}\footnote{One improvement to the sphere+disc model, not
  discussed in Dove et al.  \shortcite{dove:97b}, is the treatment of
  thermal radiation emitted by the accretion disc.  Currently, the
  temperature of the accretion disc is assumed to be $\Tbb \propto
  (R/\Rc)^{-3/4}$, where $R$ is the disc radius and $\Rc$ is the radius of
  the coronal sphere. The implicit flux of the disc (i.e., the radiation
  emitted due to viscous energy dissipation and not due to reprocessing of
  coronal radiation) is given by $F(R) = \sigma T^4(R)$, and the spectral
  shape is determined by the {\em local} disc temperature (a superposition
  of thermal Planckian distributions). Reprocessing is treated locally and
  is added to the intrinsic thermal flux from the disc to yield the total
  flux.}.  The outer radius of the cold disc is assumed to be five times
the radius of the coronal sphere ($a_r = R_d/R_c = 5$), although the
results are very insensitive to the value of this ratio for $a_r \ga 3$
\cite{dove:97b}.  The model spectra have been implemented into the data
reduction software {\it XSPEC} (version 10.0) \cite{arnaud:96a} for use in
spectral fitting. The grids of model spectra and the interpolation
routines, based on Delaunay triangularization, are available upon request.

Slab-corona models do not result in good fits, as the predicted spectra are
always much softer than the observed spectrum ($\chi^2_{\rm red} \sim 80$).
This result is consistent with our previous findings based on
non-simultaneous data \cite{dove:97b}.  The reason that the slab ADC models
always predict a spectrum softer than observed is that there is a maximum
coronal temperature for a corresponding total coronal opacity. As discussed
by Dove et al. \shortcite{dove:97b}, no self-consistent slab model can have
{\it both} a high enough temperature {\it and} a high enough opacity to
yield a Compton-$y$ parameter large enough such that the Comptonized
spectrum is as hard as that of Cyg X-1.

\begin{figure}
\centerline{\psfig{file=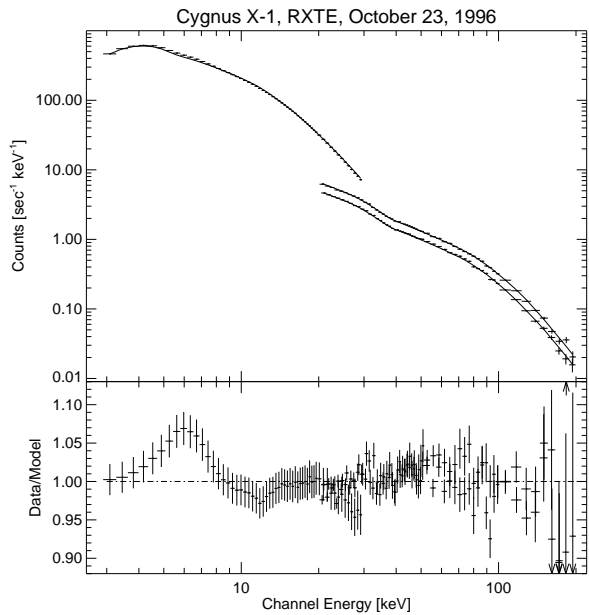,width=8.cm,angle=0.}}
\caption{Top: Best-fit sphere+disc fit to the PCA and HEXTE data. 
  Bottom: Ratio between data and model for the best-fit sphere+disc model.
  Deviations less than 2\% are within the range of uncertainty of the PCA
  response matrix.  Deviations in the $20 - 30$\,keV range are similar to those
  seen with best-fit models of Crab nebula observations (Jahoda,
  private communication).}
\label{fig:cyg-kotelp}
\end{figure}

The sphere+disc ADC model does provide a good description of the data (cf.
Table~\ref{tab1} and figure~\ref{fig:cyg-kotelp})\footnote{Note that, even
though the free parameters of each grid of models are the coronal seed
opacity and the coronal compactness parameter, for convenience we list the
corresponding {\it average} coronal temperature and {\it total} optical
depth.}.  The formal $\chi^2$-value of our best-fit model ($\chi^2_{\rm
red} = 1.55$) is larger than the values found for some of the
phenomenological models discussed in \S3.1.  Considering that our model has
only three parameters (seed optical depth, coronal compactness, overall
normalization, plus the ``hidden'' parameter $T_{\rm BB}$, which is fixed
at 150\,eV during the fitting procedure), and considering that our model is
{\it physically self-consistent}, the level of agreement with the data is
quite good.  The major disagreement between the model and the data occurs
in the $\approx 5 - 10$ keV\,range, where the residuals are as large as
7\%.  Below we discuss several ways in which slight modifications to our
model might result in improved fits to the data.

Contrary to the results of Gierli\'nski et al. \shortcite{gierlinski:97a},
we were able to fit the high-energy tail with a {\em single} Comptonizing
component.  This result might be due to the $\sim 15$\% variation of the
coronal temperature (due to the non-uniform radiation field and the
corresponding non-uniform Compton cooling rate) in our models.  We also
tried sphere+disc models in which the inner temperature of the disc is
$300$\,eV or $800$\,eV, the latter temperature roughly being consistent
with the best-fit black-body temperature found in the previous section.
However, since sphere$+$disc models calculate the flux emitted by the disc
self-consistently, both of these models predict much more of a soft-excess
than observed. This result reinforces our conclusion that the $1$\,keV
black-body component that was added to the exponentially truncated
power-law model should not be interpreted literally as disc emission.

\section{Discussion and Conclusions}\label{sec:discuss}
We have applied a variety of spectral models to an RXTE observation of
Cyg~X-1.  We find that the observed spectrum in the  $3.0 - 200$\,keV range
is well described by a power-law with a photon index $\Gamma =
\errtwo{1.45}{0.01}{0.02}$ modified by an exponential tail with an
e-folding energy $\Efold \approx 160$\,keV, and a soft excess that was
modeled as a black-body (although other broad distribution models are
capable of explaining the soft-excess) having a temperature $k\Tbb \approx
1.2$\,keV.  The measured value of $\Gamma$ is lower than those found in
previous broad-band analyses, which find a $\Gamma \sim 1.65$
\cite{gierlinski:97a,doeber:94a}. In addition, the observed strength of the
Compton reflection feature is weak, as the best-fit covering fraction,
determined by fitting the data with a power-law reflection model
\cite{magdziarz:95a}, is found to be $\approx 0.2$ when no soft excess
component is included, and is found to be $\la 0.02$ when one is
included. The latter value is considerably lower than the $f\approx 0.3$
found by Gierli\'nski et al. \shortcite{gierlinski:97a}.

Two weeks prior to our observation, Cygnus X-1 transited from a soft
(`high') state to a hard (`low') state.  For a description of the
properties the Cyg X-1 soft state, see Cui et al.  \shortcite{cui97b}.  The
proximity of our observation to this transition might have influenced our
results, leading to a harder spectrum and weaker reflection components than
what has usually been observed from this source.  However, we believe that
our best-fit parameters are also due to our fitting the entire $3 - 200$\,keV
spectral band, as discussed in section~\ref{sec:classic}.

ADC models having a slab geometry are unable to explain the observed hard
power-law and the small amount of reprocessing, consistent with the results
of Dove et al. \shortcite{dove:97b}, Gierli\'nski et al.
\shortcite{gierlinski:97a}, and Poutanen, Krolik \& Ryde
\shortcite{poutanen:97b}.  In contrast, the sphere+disc ADC models provide
a good explanation of the data, a result in agreement with
\cite{gierlinski:97a}.  Furthermore these models are {\it physically
self-consistent}.  Our sphere+disc models seem to under-produce
``reprocessing features,'' especially the Fe line, even though reprocessing
is self-consistently included within these models. For this paper, the
ratio of the accretion disc radius to the coronal sphere radius, $a$ was
set to five. As shown by Dove et al. \shortcite{dove:97b}, for $a\ga 5$,
the geometrical covering fraction of the accretion disc, as seen from the
surface of the corona, is $f_G \approx 0.30$.  If these residuals, which
represent less than 1\% of the total flux from Cyg~X-1, are due to
underestimating reprocessing, several possibilities come to mind.  First,
our model has a sharp transition between the cold (flat) disc and the
corona.  It is possible that these two regions overlap to some extent, and
thereby produce stronger reprocessing features.  Poutanen, Krolik \& Ryde
(1997) suggest that these regions overlap to a large extent during the high
state, which ended only two weeks prior to our observation.  Another
possibility is that we are observing a hot transition layer between the
disc and corona. Our disc also remains flat out to its outer edge.  If the
disc flares, which is likely due to X-ray heating, then reprocessing
features will also be enhanced.  Finally, the iron abundance in the disc of
Cyg~X-1 could be higher than the solar value that we used in our model.

Due to the data's broad energy range, the physical properties of the corona
are well constrained.  However, it is premature to claim that the
sphere+disc configuration is indeed the appropriate geometry. Other, albeit
unknown, geometries in which a small fraction of coronal radiation is
reprocessed by the disc, may also be able to describe the data. The
spectral features due to the reprocessing of coronal radiation in the cold
disc (i.e., the Fe~K$\alpha$ fluorescence line, the Compton reflection
``bump,'' and the soft-excess due to thermalization of the coronal
radiation) occur for energies in the $\sim 0.1 - 20$\,keV range, and
different geometries will predict slightly different reprocessing features.
Since we had to ignore data below $3$\,keV, we were not able to constrain
the model parameters which deal with the soft-excess due to thermal
radiation emitted by the cold disc, nor can the strength or width of the
iron line be directly measured (due to the uncertainty of the PCA response
matrix at $5.5$\,keV).  Therefore, more low energy data are needed to
further constrain ADC models for Cyg~X-1.

Additional constraints to the ADC models can come from measurements of
time-lags and the temporal coherence between several energy bands from the
source. These measurements can be compared with the time-lags and coherence
function predicted using the physical parameters of the geometry found from
spectral fitting \cite{vaughan:97a,nowak:96a}. Only a geometry in which
both the spectral and the temporal data can be explained should be
considered a valid candidate for Cyg~X-1. We are planning to use such an
approach in a forthcoming paper.

We acknowledge the help of D.~Gruber with the HEXTE data analysis, the help
of K.~Jahoda with the PCA response matrix, and useful discussions with
I.~Kreykenbohm, Ch.~Reynolds, and R.~Staubert.  This work has been financed
by NSF grants AST91-20599, AST95-29175, INT95-13899, NASA Grant NAG5-2026,
NAG5-3225, NAGS-3310, DARA grant 50\,OR\,92054, and by a travel grant to
J.W.\ from the DAAD.

\input{mntable}

\end{document}

%% file: mntable.tex
%
%
%
\begin{table*}
\caption{Results of spectral fitting to Cyg~X-1.}
\label{tab1}
\begin{tabular}{llllllllllll} 
\hline
\colhead{Model} & \colhead{$\Gamma$}          & \colhead{$\Apl$}            & 
\colhead{$\Efold$}     & \colhead{$k\Tbb$}           & \colhead{$\Abb$}         
& \colhead{$f$}            & \colhead{$k\Tc$}          &
\colhead{$\tau_{\rm T}$}      
     & \colhead{$
\Ac$}        &\colhead{$A_{\rm L}$}                   & \colhead{$\chi^2/{\rm 
dof}$} \\
                & \colhead{}                  & \colhead{}                  & 
\colhead{[keV]}        & \colhead{[keV]}             & \colhead{$10^{-2}$}      
& \colhead{}      & \colhead{keV}             & \colhead{}             
     & \colhead{}
        & \colhead{$10^{-3}$}                        &  \colhead{}              
    \\
\hline
pl              & $1.70$                      & $1.78$                      & 
\nodata                & \nodata                     & \nodata                  
& \nodata                  & \nodata                   & \nodata                
     & \nodata   
                & \nodata                & 2315/171  \\[2pt]
plexp           & $\errtwo{1.63}{0.01}{0.01}$ & $\errtwo{1.48}{0.03}{0.04}$ & 
$\errtwo{364}{28}{25}$ & \nodata                     & \nodata                  
& \nodata                  & \nodata                   & \nodata                
     & \nodata   
                & \nodata                & ~806/169 \\[2pt]
plexp+bb        & $\errtwo{1.45}{0.01}{0.02}$ & $\errtwo{0.91}{0.03}{0.04}$ & 
$\errtwo{162}{9}{8}$   & $\errtwo{1.2}{0.0}{0.1}$    & $\errtwo{2.2}{0.1}{0.2}$ 
& \nodata                  & \nodata                   & \nodata                
     & \nodata   
                & \nodata                & ~173/167 \\[2pt]
plexp+bb+g        & $\errtwo{1.45}{0.02}{0.01}$ & $\errtwo{0.91}{0.04}{0.01}$ & 
$\errtwo{164}{9}{8}$   & $\errtwo{1.1}{0.1}{0.0}$    & $\errtwo{2.1}{0.2}{0.1}$ 
& \nodata                  & \nodata                   & \nodata                
     & \nodata   
                & $\errtwo{3.3}{2.5}{2.6}$                & ~170/166 \\[2pt]
pexrav          & $\errtwo{1.71}{0.02}{0.02}$ & $\errtwo{1.67}{0.05}{0.06}$ & 
$\errtwo{924}{400}{250}$& \nodata                    & \nodata                  
& $\errtwo{0.20}{0.05}{0.05}$      & \nodata                   & \nodata                
     & \nodata   
                & \nodata                & ~752/168 \\[2pt]
pexrav+bb       & $\errtwo{1.45}{0.02}{0.01}$ & $\errtwo{0.92}{0.06}{0.05}$ & 
$\errtwo{164}{10}{10}$ & $\errtwo{1.2}{0.1}{0.1}$    & $\errtwo{2.2}{0.1}{0.2}$ 
& $\errtwo{0.01}{0.00}{0.01}$ & \nodata                & \nodata                
     & \nodata   
                & \nodata                & ~173/166 \\[2pt]
pexrav+bb+g       & $\errtwo{1.44}{0.02}{0.01}$ & $\errtwo{0.89}{0.04}{0.03}$ & 
$\errtwo{161}{9}{8}$ & $\errtwo{1.2}{0.1}{0.1}$    & $\errtwo{1.9}{0.1}{0.2}$ & 
$\errtwo{0.00}{0.02}{0.00}$ & \nodata                & \nodata                  
   & \nodata   
                & $\errtwo{3.9}{2.5}{2.6}$                & ~156/165 \\[2pt]
comptt+bb       & \nodata                     & \nodata                     & 
\nodata                & $\errtwo{1.1}{0.1}{0.1}$    & $\errtwo{1.4}{0.1}{0.1}$ 
& \nodata                  & $\errtwo{40}{1}{2}$       & 
$\errtwo{3.6}{0.1}{0.1}$    & $\errtwo{0
.47}{0.01}{0.02}$        & \nodata       & ~244/167 \\[2pt]
comptt+bb+g       & \nodata                     & \nodata                     & 
\nodata                & $\errtwo{1.0}{0.1}{0.1}$    & $\errtwo{1.3}{0.2}{0.1}$ 
& \nodata                  & $\errtwo{40}{1}{2}$       & 
$\errtwo{3.6}{0.1}{0.1}$    & $\errtwo{0
.47}{0.02}{0.01}$        & $\errtwo{5.9}{2.7}{2.6}$       & ~232/166 \\[2pt]
s+d             & \nodata                     & \nodata                     & 
\nodata                & \nodata                     & \nodata                  
& \nodata                  & $\errtwo{87}{5}{5}$ & 
$\errtwo{1.6}{0.1}{0.1}$    & $\errtwo{8.07}{0.08}{0.08}$      & \nodata         & ~263/169 \\
\hline
\end{tabular}\\
\begin{parbox}{\textwidth}
{\footnotesize $\NH$ was fixed at $6\times 10^{21}\,{\rm cm}^{-2}$.
  Uncertainties given are at the 90\% level for one interesting parameter
  ($\Delta\chi^2=2.7$).  
  pl: power-law with photon-index $\Gamma$ and
  normalization $\Apl$ (photon-flux at 1\,keV); 
  plexp: power-law with exponential cutoff 
  [$\Apl\,E^{-\Gamma}\exp(-E/\Efold)$]; 
  bb: black body with temperature $k\Tbb$; the normalization $\Abb$ is 
  defined in units of $L_{39}/R_{10}^2$ where $L_{39}$ is the luminosity in
  units of $10^{39}$\,erg/s and $R_{10}$ is the distance in units of
  10\,kpc;
  pexrav: power-law with exponentional cutoff reflected off cold matter
  (using Green's functions of Magdziarz \& Zdziarski 1995), $f$ is the
  ratio between the incident and the reflected flux; 
  comptt: Comptonization spectrum after Hua \& Titarchuk (1995) and 
  Titarchuk (1994), with a coronal temperature of $k\Tc$ and the total 
  optical depth $\tau_{\rm T}$, seed-photon temperature is 10\,eV; 
  g: Gaussian line, with energy and width fixed to 6.4 keV, 0.1 keV, 
  respectively. $A_{\rm L}$ is the amplitude of the line in total number of
  photons/cm$^2$/s in the line. $A_{\rm L}=3.3 \times 10^{-3}$ corresponds
  to an equivalent width of $\approx 45$ eV.  
  s+d: ADC model with sphere+disk geometry, {\it total} optical depth of
  the corona is $\tau_{\rm T}$, the coronal temperature is $k\Tc$, the 
   temperature profile of the cold disk is $k\Tbb(R)=150\,(R/\Rc)^{-3/4}$\,eV.}
\end{parbox}
\end{table*}